\documentclass[reprint,amsmath,amssymb]{revtex4-1}
\usepackage[utf8x]{inputenc}
\usepackage{graphicx}
\usepackage{dcolumn}
\usepackage{bm}
\usepackage{color}

\begin{document}

\title{Reconstruction of the polarization distribution of the Rice-Mele
  model}

\author{M. Yahyavi and B. Het\'enyi}

\affiliation{Department of Physics, Bilkent University 06800, Ankara, Turkey}

\begin{abstract}
We calculate the gauge invariant cumulants (and moments) associated with the
Zak phase in the Rice-Mele model.  We reconstruct the underlying probability
distribution by maximizing the information entropy and applying the moments as
constraints.  When the Wannier functions are localized within one unit cell,
the probability distribution so obtained corresponds to that of the Wannier
function.  We show that in the fully dimerized limit the magnitude of the
moments are all equal.  In this limit, if the on-site interaction is decreased
towards zero, the distribution shifts towards the midpoint of the unit cell,
but the overall shape of the distribution remains the same.  Away from this
limit, if alternate hoppings are finite, and the on-site interaction is
decreased, the distribution also shifts towards the midpoint of the unit cell,
but it does this by changing shape, by becoming asymmetric around the maximum,
as well as by shifting.  We also follow the probability distribution of the
polarization in cycles around the topologically non-trivial point of the
model.  The distribution moves across to the next unit cell, its shape
distorting considerably in the process.  If the radius of the cycle is large,
the shift of the distribution is accompanied by large variations in the
maximum.  
\end{abstract}


\maketitle

\label{sec:intro}

\section{Introduction}  

One way to derive the Berry phase~\cite{Berry84,Pancharatnam56,Xiao10} is to
form a product of scalar products between quantum states at different points
of the space of external parameters (Bargmann invariant~\cite{Bargmann64}) and
to take the continuous limit along a cyclic curve.  An
extension~\cite{Souza00,Hetenyi14} of this derivation, keeping higher order
terms, leads to gauge invariant cumulants (GIC) associated with the Berry
phase.  One is lead to ask two questions.  The GICs give information of the
distribution of what physical quantity?  Can one reconstruct the probability
distribution from the GICs?

The answer to the first question depends on the physical context in which the
Berry phase is defined.  In a crystalline solid the Berry phase (or Zak
phase~\cite{Zak89}, in this context) corresponds to the macroscopic
polarization.  Zak showed~\cite{Zak89} that the phase itself corresponds to
the expectation value of the position over a Wannier function.  For the higher
order GICs Souza, Wilkens and Martin~\cite{Souza00} showed that they only
correspond to the cumulants of the distribution of the position associated
with Wannier functions, if the Wannier functions themselves are localized
within the unit cell (non-overlapping among different unit cells).  Indeed, in
the construction of tight-binding based lattice models, one starts with a
continuum description, and assumes a localized basis of non-overlapping
Wannier functions (see for example, Ref. \cite{Essler05}).  In practice,
however, constructing such a localized basis is not trivial~\cite{Marzari97}.

The distribution of the polarization gauges the extent to which the system is
localized in the full configuration space, a criterion~\cite{Kohn64} which
distinguishes an insulator from a conductor.  The second GIC was
shown~\cite{Kudinov91,Souza00} to be proportional to the integrated frequency
dependent conductivity (sum rule).  A gauge dependent definition of the spread
(similar to the second GIC) was used to define the
maximally localized Wannier function~\cite{Marzari97}.  Also, the second
cumulant was proposed~\cite{Resta99} to distinguish conductors from
insulators.  In Ref. \cite{Hetenyi14} the simplest system with a Berry phase,
an isolated spin-$\frac{1}{2}$ particle in a magnetic field, was 
considered, and it was shown that (based on calculating the first four
cumulants) the moments of this underlying distribution are all equal.

The Zak phase was measured in Ref. \cite{Atala13} in an optical lattice setup
which corresponds to the experimental realization of the Su-Schrieffer-Heeger
(SSH) model~\cite{Su79} and its extension the Rice-Mele (RM)
model~\cite{Rice82}.  The RM model is a lattice model with an alternating
on-site potential, and hoppings with alternating strengths, depending on
whether a given bond is odd or even.  An interesting
characteristic~\cite{Vanderbilt93,Xiao10} of the RM model is its topological
behavior which manifests when an adiabatic cycle in the parameter space of the
Hamiltonian encircles the point ($\Delta=0$, $J=J'=1$).  Due to the fact that
the polarization as a function of the parameters of the Hamiltonian is not
single valued, the polarization in such a process changes by a ``polarization
quantum.''  A recent related study~\cite{Nakajima15} realized quantized
adiabatic charge pumping~\cite{Thouless83}, also in the RM model.

In this paper we calculate the leading GICs associated with the Zak phase for
the RM model.  Based on the GICs (or associated moments, GIMs) we
approximately reconstruct the distribution associated with the polarization.
The RM model is a lattice model, which implies that the underlying Wannier
functions are non-overlapping among different unit cells, and that the GICs
correspond to the distribution associated with the Wannier function.  Hence,
our reconstructed probabilities correspond to the squared modulus of the
Wannier function.  We show that in the fully dimerized limit the GIMs should
all have the same magnitude, and that the sign of odd GIMs switch sign with
respect to the direction of the polarization.  We also focus on the line of
the parameter plane where the polarization shows a line of discontinuity (see
Fig. \ref{fig:mom1}, lower panel, left inset).  We also present two model
calculations in which the evolution of the probability distribution is
followed around the topologically nontrivial point of the RM model.  As
expected, the distribution migrates to the next unit cell, although its shape
varies considerably during the cycle.

Reconstructing a probability distribution from knowledge of a finite set of
moments is an ill-posed mathematical problem which already has a long
history~\cite{Smoluchowski17}, although there has been a renewed interest in
the last decades~\cite{John07,DeSouza10}.  The scientific applications are
also quite broad; image processing~\cite{Sluzek05}, calculating magnetic
moments~\cite{Berkov00}, or molecular electronic
structure~\cite{Bandyopadhyay05}.  In our study, we opt for a reconstruction
based on maximizing the entropy~\cite{Bandyopadhyay05,Collins77,Mead84} of the
underlying probability distribution.

This paper is organized as follows.  In the next section we introduce the GICs
associated with the Zak phase.  We then discuss their connection to the
distribution associated with the Wannier functions.  In section \ref{sec:resp}
we discuss the connection of the cumulants to response functions, after which
the reconstruction procedure is presented.  In section \ref{sec:SSH} the
Su-Schrieffer-Heeger and Rice-Mele models are introduced.  Subsequently, the
behavior of the moments for the fully dimerized limit is studied.  Section
\ref{sec:results} contains our results and analysis before concluding our
work.

\section{Gauge invariant cumulants associated with the Zak phase}
\label{sec:cml}

Consider a one-dimensional system whose Hamiltonian which is periodic in $L$.
We take Bloch functions parametrized by the crystal momentum, $\Psi_0(K)$ on a
grid of $M$ points $K_I = 2\pi I /(M L) - \pi/L$, with $I=0,...,M-1$.  The Zak
phase can be derived from a product of the form
\begin{equation}
\label{eqn:phi_dscrt}
\phi_{Zak} = \mbox{Im} \ln \prod_{I=0}^{M-1} \langle \Psi_0(K_I)|\Psi_0(K_{I+1})\rangle,
\end{equation}
by taking the continuous limit ($M\rightarrow \infty$).  The product in
Eq. (\ref{eqn:phi_dscrt}) is known as the Bargmann
invariant~\cite{Bargmann64}.  We will derive the Zak phase, as well as the
associated gauge invariant cumulants (GIC).  We start by equating the product
in Eq. (\ref{eqn:phi_dscrt}) to a cumulant expansion,
\begin{equation}
\label{eqn:cum_exp} 
\left[\prod_{I=0}^{M-1} \langle
  \Psi_0(K_I)|\Psi_0(K_{I+1})\rangle\right]^{\Delta K} = 
\exp\left(  \sum_{n=1}^\infty \frac{(i \Delta K)^n}{n!}\tilde{C}_n\right),
\end{equation}
with $\Delta K = 2 \pi/M$.  We now expand both sides and equate like powers of
$\Delta K$ term-by-term, mindful of the fact that the left-hand side includes
a product over $I$.  For example, the first-order term will be
\begin{equation}
\tilde{C}_1 = i \sum_{I=0}^{M-1}\Delta K
\gamma_1(K_I)
\end{equation}
the second will be
\begin{equation}
\tilde{C}_2 = - \sum_{I=0}^{M-1}\Delta K [\gamma_2(K_I) - \gamma_1(K_I)^2]
\end{equation}
with $\gamma_i(K) = \langle \Psi_0(K) | \partial^i_K |
\Psi_0(K)\rangle$.  Straightforward algebra and taking the continuous
limit ($\Delta K \rightarrow 0$, $M\rightarrow \infty$)
gives up to the fourth order term,
\begin{eqnarray}
\label{eqn:cmlnts}
C_1 &=& i \frac{L}{2\pi} \int_{-\frac{\pi}{L}}^{\frac{\pi}{L}} d K \gamma_1 \\ \nonumber 
C_2 &=& -\frac{L}{2\pi} \int_{-\frac{\pi}{L}}^{\frac{\pi}{L}} d K [\gamma_2 - \gamma_1^2] \\ \nonumber
C_3 &=& -i \frac{L}{2\pi} \int_{-\frac{\pi}{L}}^{\frac{\pi}{L}} d K [\gamma_3 -3 \gamma_2 \gamma_1+ 2\gamma_1^3] \\ \nonumber 
C_4 &=& \frac{L}{2\pi} \int_{-\frac{\pi}{L}}^{\frac{\pi}{L}} d K [\gamma_4 -3 \gamma_2^2 -4\gamma_3\gamma_1 + 12 \gamma_1^2\gamma_2 -6\gamma_1^4]
\end{eqnarray}
The quantities $C_n$ in Eq. (\ref{eqn:cmlnts}) are the GICs associated with
the Zak phase (the Zak phase itself being equal to $C_1$).  The difference
between $\tilde{C}_i$ and $C_i$ is the multiplicative factor $L/2\pi$, which
is also how the phase is defined by Zak~\cite{Zak89}.  This assures that the
first moment corresponds to the average position associated with square
modulus of the Wannier function (Eq. (10) in Ref. ~\cite{Zak89}).  When the
  underlying probability distribution is well defined the associated moments
  can be defined based on the cumulants.  Following this standard procedure we
  also define a set of moments.  For the first four moments the expressions
  are
\begin{eqnarray}
\label{eqn:muC}
\mu_C^{(1)} &=& C_1 \\ \nonumber 
\mu_C^{(2)} &=& C_2 + C_1^2 \\ \nonumber
\mu_C^{(3)} &=& C_3 + 3 C_2 C_1 + C_1^3 \\ \nonumber 
\mu_C^{(4)} &=& C_4 + 4 C_3 C_1 + 3 C_2^2 + 6 C_2 C_1^2 + C_1^4.
\end{eqnarray}
As discussed below, when the Wannier functions of a particular model are
localized within the unit cell, these moments correspond to the moments of the
polarization, alternatively, to the distribution of the Wannier functions
themselves.

We remark that in general, the Berry phase is a physically well-defined
observable, which is thought not to correspond to an operator acting on the
Hilbert space.  The Zak phase, however, is known to correspond to the total
position, and is the basic quantity in expressing the polarization in the
modern theory~\cite{King-Smith93,Resta94,Resta98}.

\section{Connection to the distribution of Wannier centers}

Cumulants of the type described in the previous section appear in the theory
of polarization~\cite{Souza00}.  In this section we connect the cumulants to
the distribution of Wannier centers.  We consider a typical term contributing
to cumulant $C_M$, which can be written in the form
\begin{equation}
\label{eqn:C_m}
C_{M,\alpha} = \frac{L}{2 \pi}
\int_{-\frac{\pi}{L}}^{\frac{\pi}{L}} dK \prod_{i=1}^d \langle u_{nK} | \partial^{m_i}_K| u_{nK} \rangle,
\end{equation}
where $\sum_{i=1}^d m_i = M$ and where we have used the periodic Bloch
functions $u_{nK}(x)$ as a basis.  The periodic Bloch functions can be written
in terms of Wannier functions,
\begin{equation}
u_{nK}(x) =  \sum_{p=-\infty}^\infty \exp(iK(p L - x)) a_n(x - p L),
\end{equation}
where $a_n(x)$ denote the Wannier functions.  With this
  definition it holds that
\begin{equation}
\frac{L}{2\pi} \int_{-\pi/L}^{\pi/L} d K \int_0^L d x |u_{nK}(x)|^2 =
\int_{-\infty}^{\infty} d x |a_n(x)| = 1.
\end{equation}

We can rewrite a scalar
product appearing in Eq. (\ref{eqn:C_m}) as 
\begin{eqnarray}
\nonumber 
\langle u_{nK} | \partial^{m}_K| u_{nK} \rangle =  \sum_{ \Delta p= - \infty}^\infty \exp(-iK \Delta p L )
\int_{-\infty}^\infty dx  \\ a_n^*(x - \Delta p L) (-i x)^m a_n(x). \hspace{2cm}
\label{eqn:scalarprod}
\end{eqnarray}
Substituting Eq. (\ref{eqn:scalarprod}) $C_{M,\alpha}$ and integrating in
$K$ results in
\begin{eqnarray}
C_{M.\alpha} = 
\sum_{\Delta p_1 = -\infty}^\infty 
...
\sum_{\Delta p_d = -\infty}^\infty \delta[\Delta P , 0] \hspace{1cm}\\ \hspace{.5cm}
\prod_{j=1}^d 
\left \{ \int_{-\infty}^\infty dx_j (-i x_j)^{m_j} 
a_n^*(x_j - \Delta p_j L)  a_n(x_j) \right \}
, \nonumber
\end{eqnarray}
where $\Delta P = \sum_{j=1}^d \Delta p_j$ and $\delta[\Delta P , 0]$ is a
Kronecker delta.  

We note that if the Wannier functions are localized in one unit cell,
then the summation in the scalar product of Eq. (\ref{eqn:scalarprod})
will be restricted to the term $\Delta p = 0$.  In this case, the
cumulants $C_M$ will correspond to those of the Wannier centers.

\section{Relation to response functions}
\label{sec:resp}

The second GIC associated with the polarization gives a sum rule for the
frequency-dependent conductivity.  This was shown for a finite system by
Kudinov~\cite{Kudinov91}, and the derivation was extended to periodic systems
by Souza, Wilkens, and Martin~\cite{Souza00}, by replacing the ordinary matrix
elements of the total position operator by their counterparts valid in the
crystalline case.  Their result is
\begin{equation}
  C_2 = \frac{\hbar}{\pi q_e^2 n_0} \int \frac{d \omega}{\omega} \bar{\sigma}(\omega),
\end{equation}
where $q_e$ denotes the charge, $n_0$ the density, and
$\bar{\sigma}(\omega)=(V/8\pi^3)\int d{\bf k} \sigma^{\bf k}(\omega)$.

For an insulating (gapped) system one can show that the second cumulant
provides an upper bound for the dielectric susceptibility, $\chi$.  This was
shown by Baeriswyl~\cite{Baeriswyl00} for an open system.  This derivation is
also easily extended to periodic systems by the appropriate replacement of the
total position matrix elements, resulting in,
\begin{equation}
  \label{eqn:chi}
  \chi \leq \frac{2q_e}{V \Delta_g} C_2.
\end{equation}
In this equation $\Delta_g$ denotes the gap, $V$ denotes the volume of
the system.

For higher order cumulants, the derivation of relations such as
Eq. ($\ref{eqn:chi}$) are not possible.  However, in the classical limit,
the cumulants correspond exactly to the response functions of their
respective order ($C_2$ gives $\chi$, $C_3$ gives the first non-linear
response function, etc.).

\section{Reconstruction of the probability distribution}  

If the Wannier functions can be assumed to be localized within a unit cell,
the moments calculated based on the GICs correspond to the actual moments
associated with the Wannier orbitals.  If all the moments are
  known, the full probability distribution can be reconstructed.  However, in
  practice, usually only a finite number of cumulants are available.  In this
  case the cumulants can be used as constraints to improve the form of the
  probability distribution.  The first and second cumulants give the average
  and the variance, and if only these two are available, the best guess for
  the probability distribution is a Gaussian.  Higher order cumulants refine
  this guess.  The third cumulant (skewness) provides information about the
  asymmetry of the distribution around the mean, while the fourth order one,
  (kurtosis) represents how sharp the maximum of the distribution is
  approached from either side.

Below we calculate the GICs of the Rice-Mele model, which is a lattice model
(in other words, the Wannier functions are completely localized on particular
sites), and approximately reconstruct the probability distribution of the
polarization.  Our reconstruction is
based~\cite{Bandyopadhyay05,Collins77,Mead84} on maximizing the information
entropy under the constraints provided by the moments calculated.  The
  expression for the entropy we use is
\begin{equation}
\label{eqn:SPx}
S[P(x)] = - \int d x P(x) \ln P(x),
\end{equation}
minimized as a functional of $P(x)$ under the constraints
\begin{equation}
\mu_P^{(k)} = \int d x P(x) x^k,
\end{equation}
as well as the constraint that $P(x)$ is normalized.  The functional
minimization of Eq. (\ref{eqn:SPx}) under the constraints results in the
functional differential equation
\begin{equation}
\label{eqn:Sopt}
\frac{\delta}{\delta P(x)} \left[ S[P(x)]
- \sum_k A_k (\mu_P^{(k)} - \mu_C^{(k)})
 \right] = 0, 
\end{equation}
where $\mu_C^{(k)}$ are the moments obtained from the cumulants of the Berry
phase (see Eq. (\ref{eqn:muC})), and $A_k$ are Lagrange multipliers.  The
solution of Eq. (\ref{eqn:Sopt}) is
\begin{equation}
P(x) = C\exp\left( - \sum_k A_k x^k\right),
\end{equation}
where $C$ is the normalization constant.  We determine the constants $A_k$ by
numerically minimizing the quantity
\begin{equation}
\label{eqn:chisquared}
\chi^2 = \sum_k (\mu_P^{(k)}-\mu_C^{(k)})^2,
\end{equation}
as a function of $A_k$.  As our initial guess in all cases studied below, we
take the Gaussian distribution defined by the first two cumulants obtained for
the particular case.  The minimization procedure we applied is the simulated
annealing technique~\cite{Kirkpatrick83}.  Below our reconstructions are based
on calculating the first six GIMs in all cases.

\section{Su-Schrieffer-Heeger and Rice-Mele models}  
\label{sec:SSH}

The SSH model was first introduced~\cite{Su79} to understand the properties of
one-dimensional polyacetylene.  The RM model is an extension of the SSH model,
it includes an additional term, consisting of an alternating on-site
potential, added in order to extend the SSH model to diatomic polymers.  In
recent decades it has been studied extensively due to the wealth of
interesting physical phenomena it displays: topological soliton excitation,
fractional charge, and non-trivial edge
states\cite{Takayama80,Su80,Jackiw76,Heeger88,Ruostekoski02,Li14}.  It was
also realized as a system of cold atoms trapped in an optical lattice in one
dimension recently~\cite{Atala13}.  The Berry phase in the RM model was
studied by Vanderbilt and King-Smith~\cite{Vanderbilt93}.  In that study the
point of the model in parameter space of the model which is metallic (and
which is responsible for the topologically nontrivial behavior) was encircled
in parameter space.  This leads to the increase of $C_1$ (the Berry phase, or
the polarization) by one polarization quantum, consistent with the
quantization of charge transport~\cite{Thouless83,King-Smith93}.

\begin{figure}[ht]
 \centering
 \includegraphics[width=\linewidth]{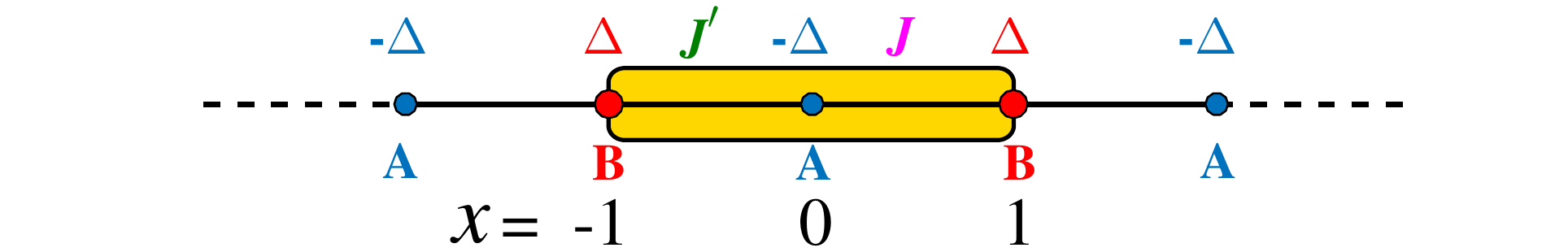}
 \caption{(Color online.) Schematic representation of the Rice-Mele model.
   $\Delta$ represents the on-site potential, $A$ and $B$ refer to the
   different sublattices.  $J$ and $J'$ are the alternating hoppings.  The
   unit cell is indicated in yellow.  The $x$ label corresponds to
   localization within the unit cell ($-1<x<1$).  The variable $x$
   is continuous, below, in our subsequent calculations, the probability
   distribution will be shown as a function of $x$.  The unit of $x$ is the lattice constant.}
 \label{fig:model}
\end{figure}

The hopping part of the SSH Hamiltonian reads:
\begin{equation}
\hat{H}_{SSH} = -J \sum_{i=1}^{N/2}  c_{i,A}^\dagger c_{i,B}
               -J' \sum_{i=1}^{N/2}  c_{i,B}^\dagger c_{i+1,A} + \mbox{H.c.},
\end{equation}
where $N$ denotes the number of sites, the on-site potential has the form
\begin{equation}
\hat{H}_{\Delta} = -\Delta \sum_{i=1}^{N/2}  c_{i,A}^\dagger c_{i,A}
               + \Delta \sum_{i=1}^{N/2}  c_{i,B}^\dagger c_{i,B}.
\end{equation}
The model is shown schematically in Fig. \ref{fig:model}.  This figure shows
the one-dimensional lattice, including sublattices, the alternating hoppings,
and the on-site potential.  The unit cell is indicated in shaded yellow.  Also
shown is the continuous variable $x$, which runs from $-\infty$ to $\infty$,
and will serve as the axis for the reconstructed probability distributions of
the polarization calculated below.

The hoppings can also be expressed in terms of the average hopping $t$ and the
deviation $\delta$ as
\begin{equation}
\label{eqn:tdelta}
J = \frac{t}{2} + \frac{\delta}{2}, J' = \frac{t}{2} - \frac{\delta}{2}.
\end{equation}
The total Hamiltonian we consider is
\begin{equation}
\hat{H} = \hat{H}_{SSH} + \hat{H}_{\Delta}.
\end{equation}
The parameters $J$ and $J'$ are hopping parameters corresponding to hopping
along alternating bonds.  We take the lattice constant to be unity (the unit
cell is two lattice constants).  The parameter $\Delta$ denotes the on-site
potential, whose sign alternates from site to site.  This model is metallic
for $J=J'$ and $\Delta=0$ but is insulating for all other values of the
parameters.  In reciprocal space this Hamiltonian becomes
\begin{equation}
\hat{H} = \sum_k \left( 
\begin{array}{cc} \Delta & -\rho_k \\ 
-\rho_k^* & -\Delta, 
\end{array}
\right)
\end{equation}
where
\begin{equation}
\rho_k = J e^{ik} + J' e^{-ik}.
\end{equation}
At a particular value of $k$ we can write the eigenstate for the lower band as
\begin{equation}
\left( 
\begin{array}{c} \alpha_k \\ 
\beta_k
\end{array}
\right) = 
\left( 
\begin{array}{c} \sin\left(\frac{\theta_k}{2}\right) \\ 
e^{-i \phi_k}\cos\left(\frac{\theta_k}{2}\right)
\end{array}
\right),
\end{equation}
where
\begin{eqnarray}
\theta_k &=& \arctan \left( \frac{|\rho_k|}{\Delta}\right) \\
\phi_k &=& \arctan \left( \frac{(J-J')\sin(k)}{(J+J')\cos(k)}\right). \nonumber
\end{eqnarray}
The cumulants can now be written in terms of the eigenstates.  For example,
\begin{equation}
C_1 = \frac{i}{\pi} \int_{-\pi/2}^{\pi/2} dk ( \alpha_k^* \partial_k \alpha_k + \beta_k^*
\partial_k \beta_k ),
\end{equation}
and the other cumulants can be constructed accordingly (note that the unit
cell is $L=2$).

\section{Fully dimerized limit}

\label{sec:FDL}

Here we show that in the fully dimerized limit the GIMs should all have the
same magnitude.  In Ref. \cite{Hetenyi14} we pointed out that the Berry phase
can be related to an observable $\hat{O}$ fixed by requiring that
\begin{equation}
\label{eqn:cnd}
\partial_K H(K) = i [H(K),\hat{O}].
\end{equation}
This definition does not uniquely fix the operator $\hat{O}$.  For example,
for the magnetic field example the matrix $\sigma_z/2$ or $(\sigma_z +I)/2$
both satisfy Eq. (\ref{eqn:cnd}).  This arbitrariness causes a shift in the
first cumulant.  However, only the operator $(\sigma_z+I)/2$ will give a
distribution in which all moments are equal, since this matrix has the form
\begin{equation}
(\sigma_z+I)/2 = \left( 
\begin{array}{cc} 1 & 0  \\ 
0 & 0
\end{array}
\right),
\end{equation}
and is equal to itself when raised to any power.

In the case of the RM model we first write the Hamiltonian with the parameter
$K$ explicitly as
\begin{eqnarray}
\hat{H}(K) &= -J \exp(i K) \sum_{j=1}^{L/2}  c_{j,A}^\dagger c_{j,B} + \mbox{H.c.}\\ 
& \nonumber
         -J' \exp(i K) \sum_{j=1}^{L/2}  c_{j,B}^\dagger c_{j+1,A}  + \mbox{H.c.} + \hat{H}_{\Delta}. \nonumber
\end{eqnarray}
The operator $\partial_K \hat{H}(K)$ is the current,
\begin{eqnarray}
\partial_K \hat{H}(K) &= -i J \exp(i K) \sum_{j=1}^{L/2}  c_{j,A}^\dagger c_{j,B} + \mbox{H.c.}\\ 
& \nonumber
         -i J' \exp(i K) \sum_{j=1}^{L/2}  c_{j,B}^\dagger c_{j+1,A}  + \mbox{H.c.}. \nonumber
\end{eqnarray}
We now write a form for the operator $\hat{O}$ as
\begin{equation}
\hat{O} = \sum_{j=1}^{L/2} x_j c_{j,A}^\dagger c_{j,A} + 
y_j c_{j,B}^\dagger c_{j,B}.
\end{equation}
Evaluating the commutator gives
\begin{eqnarray}
i [\hat{H}(K),\hat{O}] = i \sum_{j=1}^{L/2}(y_j - x_j) J \exp(iK) c_{j,A}^\dagger c_{j,B} + \mbox{H.c.} \nonumber \\  
i \sum_{j=1}^{L/2}(x_{j+1} - y_j) J' \exp(iK) c_{j+1,A}^\dagger c_{j,B} + \mbox{H.c.} \hspace{1cm}
\end{eqnarray}
For the case $J'=0$ we can chose $x_j=0$ and $y_j=1$, so that
$i[\hat{H}(K),\hat{O}]$ corresponds to the current.  This is not the
only choice, but with this choice the operator $\hat{O}$ when written
in $k$-space corresponds to
\begin{equation}
 \hat{O} = \sum_k ( c^\dagger_{k,A}  c^\dagger_{k,B} )
\left( 
\begin{array}{cc} 0 & 0  \\ 
0 & 1
\end{array}
\right)
\left( 
\begin{array}{c} c_{k,A}  \\ 
c_{k,B} 
\end{array}
\right),
\end{equation}
which gives equal moments.  Clearly, the choice for the spatial coefficients
$x_j$ and $y_j$ are due to the fact that in this case the system consists of a
set of independent dipoles.  When $J$ is taken to zero, and $J'$ kept finite,
then the appropriate choice to fix $\hat{O}$ is $x_j=0$ and $y_j=-1$.  If
instead the sign of $\Delta$ is changed $\hat{O}$ is again defined by the
$x_j=0$ and $y_j=-1$.  These results are clearly due to the reversal of the
direction of the dipole moment within the unit cell.  The results presented in
Fig. \ref{fig:mom01} corroborate our derivation.

\section{Results and Analysis}  
\label{sec:results}

We first look at the system with $J'=0$.  In this case, the band structure of
the system is simply two flat lines in the Brillouin zone.  The system can be
thought of as a simple two-state system.  We calculated the first four GICs,
from which we obtained the corresponding GIMs.  The results are shown in the
uppermost panel of Fig. \ref{fig:mom01}.  The moments as a function of
$\Delta/J$ all fall on the same curve in this case.  If the hopping parameters
$J$ and $J'$ are switched (not shown), the sign of the odd moments changes,
the even moments remain the same.  These results are in accordance with
section \ref{sec:FDL}.

Fig. \ref{fig:mom01} also shows the cumulants for different ratios;
$J'/J=0.3, 0.5, 0.7$.  The deviation of the cumulants from one another is more
pronounced, and increases with an increase of $J'/J$.  However, the moments
become equal for any $J'/J$ when $\Delta \rightarrow \pm \infty $.  In this
case also, the system becomes an independent array of two state systems.  The
band energies in all these cases vary continuously with $k$ across the
Brillouin zone.  

The results for the case $J'/J=1$ are also shown separately in Fig.
\ref{fig:mom1}, as well as the limits $J' \rightarrow_\pm J$.  For finite
$\Delta/J$ the odd cumulants are zero, indicating an even probability
distribution.  The ratio of the second and fourth cumulants rule out a
Gaussian.  As $\Delta/J \rightarrow 0$ a discontinuity in the slope of the
band develops.  In this case, the cumulants $C_2$ and $C_4$ diverge.
The lower panel in this figure shows what happens when $J'$ is
  close to $J$ (bigger or smaller) but the two are not quite equal ($J' = J +
  \epsilon$, $\epsilon$ a small number).  We see that in this case the first
  moment is one or minus one, depending on the sign of $\epsilon$, and zero is
  not approached as $\epsilon \rightarrow 0$ from either side.  The left inset
  in the lower panel shows the behavior of the first moment on the
  $\Delta-\delta$ plane, indicating a discontinuity along the line $\Delta<0,
  \delta = 0$ (the well-known result of Vanderbilt and
  King-Smith~\cite{Xiao10,Vanderbilt93}).  The moments and cumulants we find
  are consistent with the behavior shown in the left inset of the lower panel
  of the figure. 

In Fig. \ref{fig:prob} we show examples of reconstructed probability
distributions for $J'/J=0.0, 0.3, 0.5, 0.7$, in each case for several values
of $\Delta/J$.  $\chi^2$ (defined in Eq. (\ref{eqn:chisquared})) is tabulated
in the appendix (Table \ref{tab:chi2}).  The most localized example ($J'/J =
0$ and $\Delta/J=-2$) shows a sharp peak around $x=1$; as $\Delta$
decreases the curves shift to the left and spread out, but their shape is
always very similar (for smaller values of $\Delta/J$ this is emphasized in
the inset).  The maximum of the probability distribution is
  always between zero and one.  These curves are all cases for which all the
moments are equal.  As the alternate hoppings ($J'$) are turned on, the
shifting occurs in a qualitatively different manner.  Initially ($\Delta/J=-2$
in all cases) the curves are centered very near $x=1$.  $\Delta/J=-2$ is
for most cases well in the region where the moments are equal.  As $\Delta$
decreases, the distributions shift, but they do this by becoming asymmetric
about their mean, with the density increasing on the side left of the maxima
of the distributions.  The shape of the distributions changes considerably.
This is clearly due to the fact that in these latter cases the moments vary as
$\Delta$ is varied, and they are not all equal.  The maxima for the cases for
which $J'/J \neq 0$ shift much less as $\Delta/J$ is varied.  When $\Delta/J$
changes sign (results not shown), the polarization becomes centered around
$x=0$ end of the unit cell and the probability distributions are
reflections of the ones shown in Fig. \ref{fig:prob} across $x = 1/2$.

The probability distributions for the case $J'=J=1$ are also shown separately
in Fig. \ref{fig:prob1} (with $\chi^2$ tabulated in Table \ref{tab:chi2}), as
well as the case $J'$ close to $J$.  All of the $J'=J$ distributions are
symmetric around the origin.  As $\Delta/J \rightarrow 0$ the distribution
broadens, and it is clear that a conducting phase is
approached~\cite{Resta99}.  If $\Delta<0$ then the
  polarizations are localized near $x = \pm1$ depending on whether $J'$ is
  smaller or larger than $J$.  This is consistent with Fig. \ref{fig:mom1}.

In Figs. \ref{fig:CircleR1} and \ref{fig:CircleR02} we show the evolution of
the reconstructed probability distributions along two cyclic paths which
encircle the topologically non-trivial point of the RM model, one with radius
unity, the other with radius $0.2$ in the $\Delta/t$, $\delta/t$ plane.
In these calculations the parametrization was different from
  the previous ones, here $t$ was set to unity, rather than $J$ (see
  Eq. (\ref{eqn:tdelta})).  For the points $A^*$, $B$, ... in
  Figs. \ref{fig:CircleR1} and \ref{fig:CircleR02} the values of $\Delta/J$
  and $J'/J$ are shown in Table \ref{tab:chi2_circles}.  The upper panel in
both figures show the evolution of the different GIMs(GICs).  The even moments
are single-valued, the odd ones are not.  This follows from gauge invariance
properties of the cumulants (Eq. (\ref{eqn:cmlnts})).  The first cumulant is
only gauge invariant modulo $2 \pi$ times an
integer~\cite{Xiao10,Vanderbilt93}, the others do not change at all due to a
gauge transformation.  The odd GIMs depend on combinations of the GICs which
involve odd combinations of the cumulants, therefore they are not multivalued
in general.  In both sets of figures (\ref{fig:CircleR1} and
  \ref{fig:CircleR02}) the points $A^*$ are not exactly on the $\phi = -\pi/2$
  axis, but instead we numerically realize the limit $\phi = \lim_{\delta \Phi
    \rightarrow 0^+} (-\pi/2 + \delta \Phi)$.  In the actual calculation we
  took $\delta \Phi = 2 \pi/1000$.  Also, the point $\phi = -\pi/2$ or $\phi =
  3\pi/2$ is excluded from the curves shown in the upper panels of the two
  figures.

The example with radius unity (Fig. \ref{fig:CircleR1}) remains mostly in the
fully dimerized limit, as can be seen in the upper panel of the figure.  The
odd moments and even moments are always equal.  Except for a small region near
$\phi/\pi = 0.5$ the absolute values of the moments are equal.  The lower
panel shows the evolution of the probability distribution along the path.
Starting from a relatively sharp distribution localized near $x=1$, the
maximum moves to the left.  Before reaching half the unit
cell, the distribution spreads.  After passing through the midpoint the
system, where the maximum is the smallest, the distribution begins to localize
again until $x=0$. From there this tendency is repeated.  Indeed, the
distribution ends up at $x=-1$ at the end of the process: the Wannier
function ``walked'' to an equivalent position in the next unit cell.  For the
case of the smaller radius (0.2, Fig. \ref{fig:CircleR1}) the initial
distribution is broader, and as the cycle is traversed, the maximum of the
distribution oscillates with a smaller amplitude, but the ``walking'' to a new
equivalent position still occurs.

In both Figs. \ref{fig:CircleR1} and \ref{fig:CircleR02} it is clear that the
odd moments do not correspond to single-valued functions.  The values of the
odd moments depend on whether we approach the original point from which the
cycle begins ($\delta = 0, \Delta<0$) from the left or the right.  At the same
time, the probability distributions for some cases with $\delta = 0, \Delta<0$
are shown in Fig. \ref{fig:prob1}; they are centered around zero and they
spread as $\Delta/J$ is decreased.  This suggests the limiting cases from
either direction give different results from the result for fixing the
Hamiltonian parameters such that $\delta = 0,\Delta<0$.

\section{Conclusion}

We studied the gauge invariant cumulants associated with the Zak phase.  We
have shown that for localized Wannier functions they correspond to the
cumulants of the Wannier centers.  They are also related to the dielectric
response functions of a given system.  We calculated the cumulants for the
Rice-Mele model.  In the limit of isolated dimers, all the moments (extracted
from the gauge invariant cumulants) are equal.  This can be justified for this
case by constructing the operator which corresponds to the Berry phase
explicitly.  Deviations from this behavior come about when the hopping
parameters are both finite.  For a system with equal hopping parameters the
odd cumulants vanish.  We have also reconstructed the full probability
distribution of the polarization based on the gauge invariant cumulants and
have studied how they evolve as functions of different parameters of the
Hamiltonian.  In particular we calculated the evolution of the distribution
around the topologically non-trivial point of the model.  We anticipate that
detailed experimental measurements can also provide a probability distribution
of the polarization for comparison with our predictions.

\section*{Acknowledgments}

The authors acknowledge financial support from the Turkish agency for basic
research (T\"UBITAK, grant no. 113F334).   We also thank L. G. M. de Souza for
helpful discussions on the topic of probability reconstruction from moments.

\section*{Appendix}

\begin{figure}[ht]
 \centering
 \includegraphics[width=0.9\linewidth]{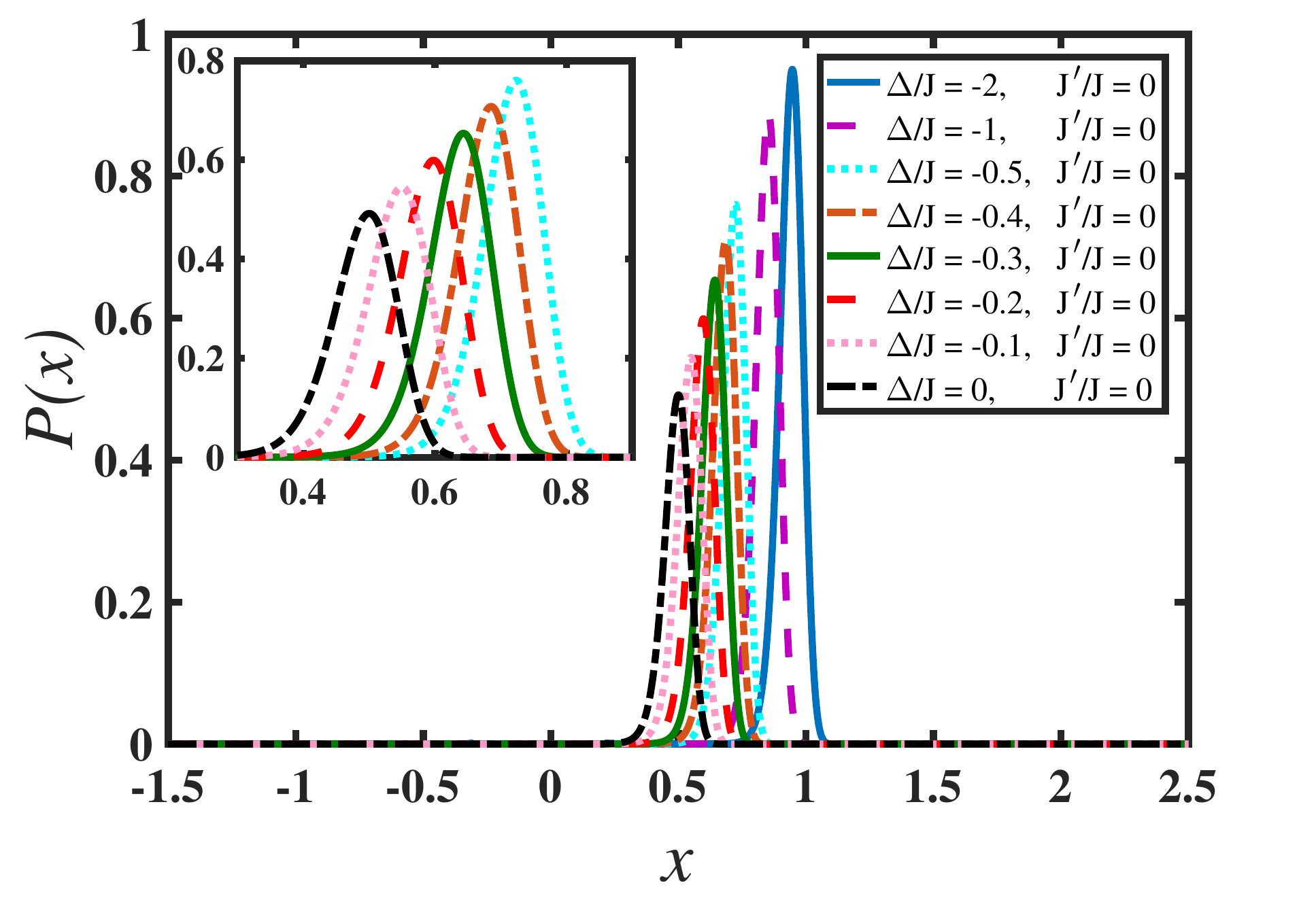}
 \includegraphics[width=0.9\linewidth]{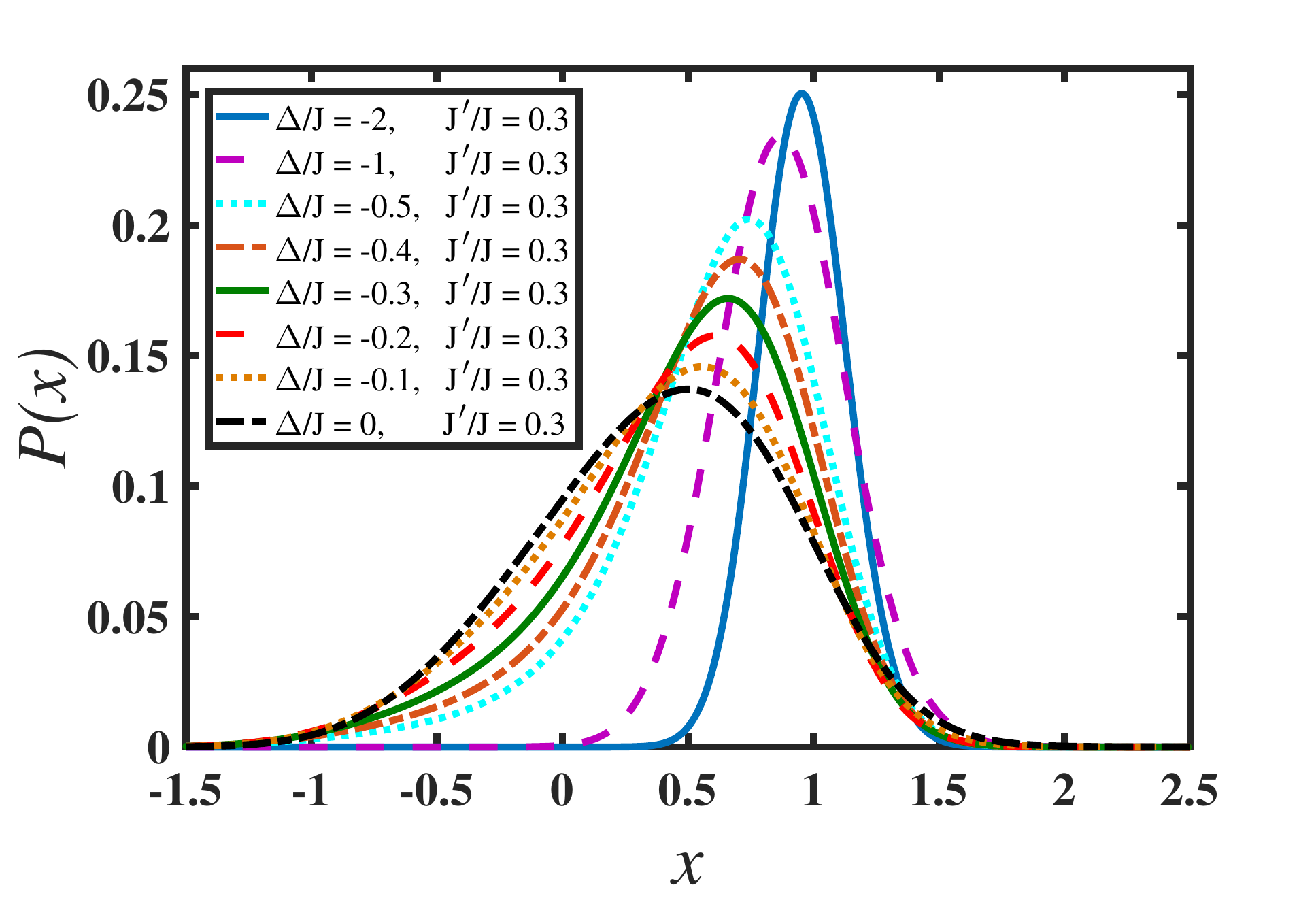}
 \includegraphics[width=0.9\linewidth]{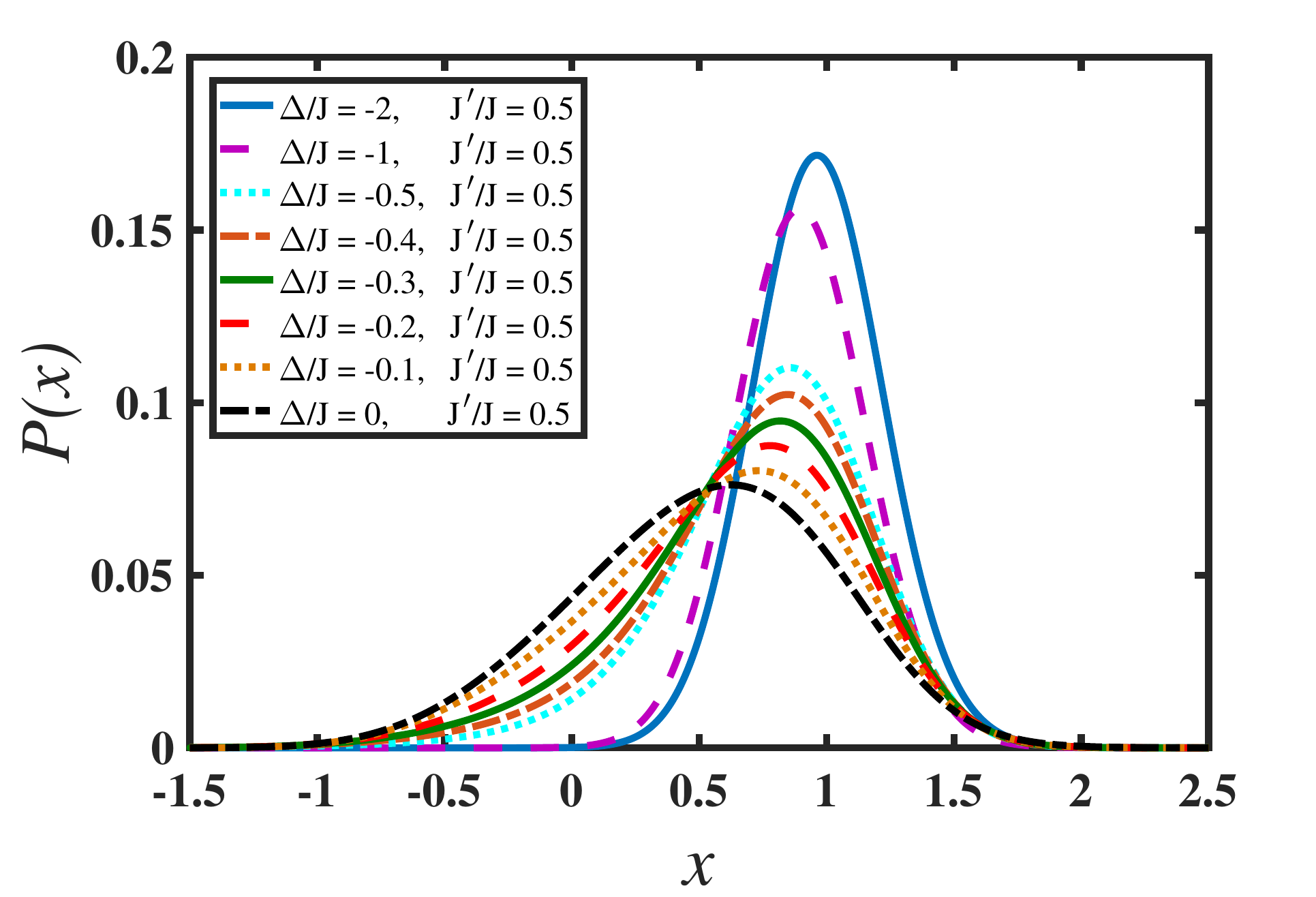}
 \includegraphics[width=0.9\linewidth]{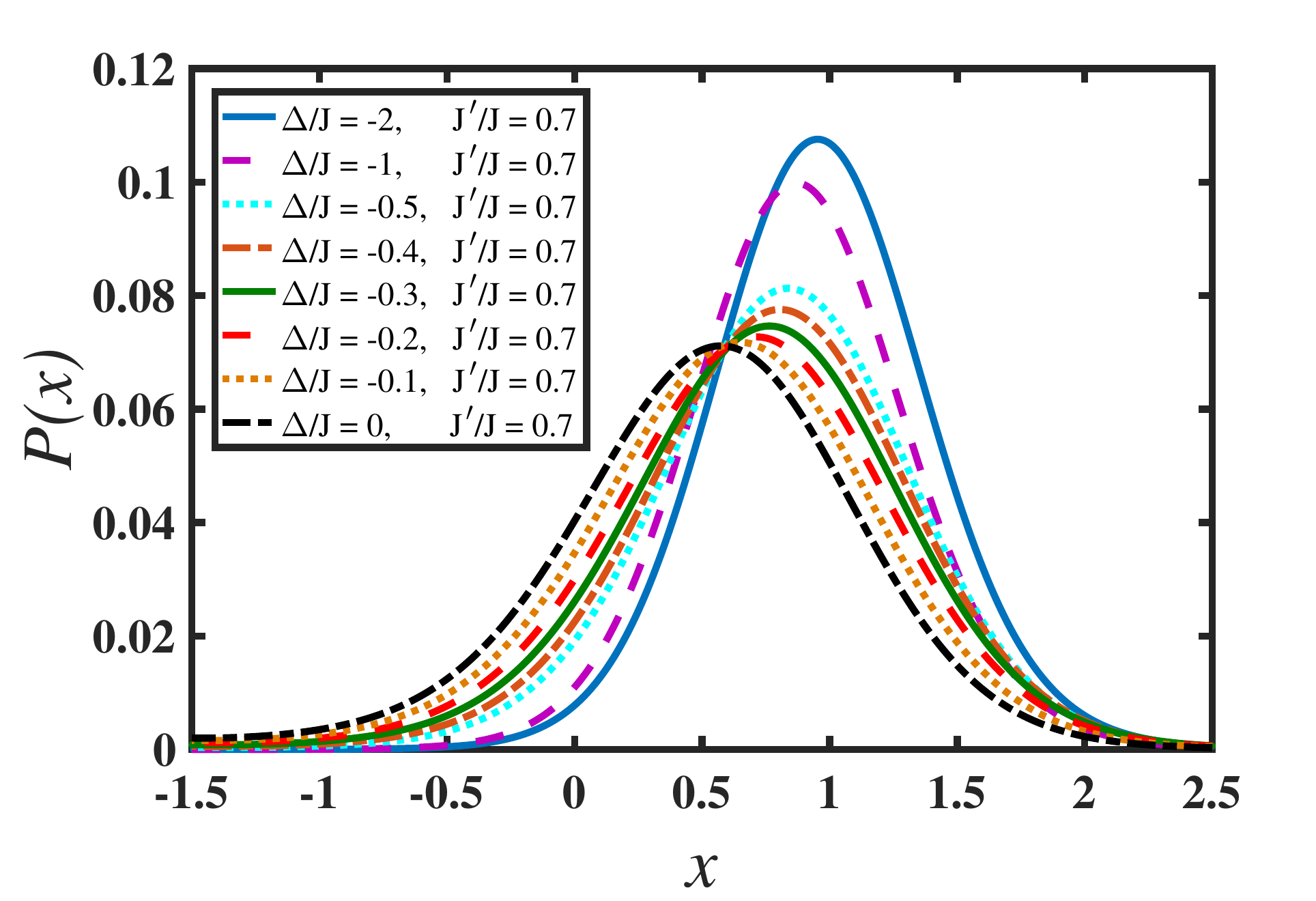}
 \caption{(Color online.)  Normalized probability distribution of the
   polarization for different parameters of the Rice-Mele Hamiltonian.  In
   these calculations $J=1$.  The unit of length in these figures is the
   lattice constant.  Different values of $\Delta/J$ are shown for $J'/J=0.0,
   0.3, 0.5, 0.7$.  In the topmost panel ($J'/J=0$) the inset shows the
   distribution for the cases $\Delta/J = 0.0,-0.1,-0.2,-0.3,-0.4,-0.5$)}
 \label{fig:prob}
\end{figure}

\begin{figure}[ht]
 \centering
 \includegraphics[width=\linewidth]{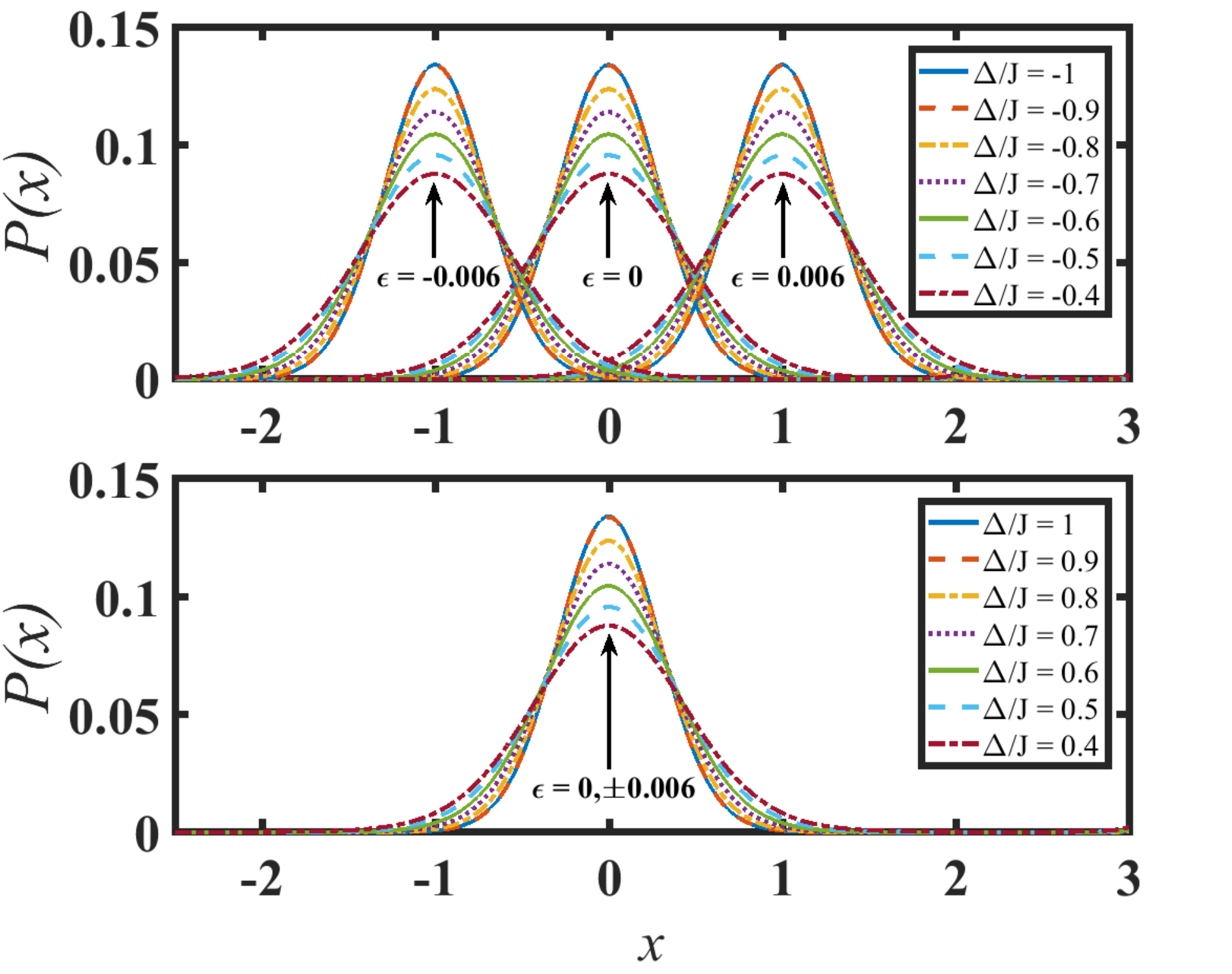}
 \caption{(Color online.)  Normalized probability distribution of the
   polarization for cases $J'=J$ and $J' = J \pm \epsilon$ ($\epsilon =
   0.006$).  In these calculations $J=1$.  The unit of length is the lattice
   constant.  Different values of $\Delta/J$ are shown.  Upper panel(lower
   panel): $\Delta /J < 0$ ($\Delta /J > 0$).}
 \label{fig:prob1}
\end{figure}

In Table \ref{tab:chi2} values of the negative base ten logarithm of $\chi^2$
rounded down to the first digit (defined in Eq. (\ref{eqn:chisquared})) is
shown for the reconstructed probabilities in Figs. \ref{fig:prob} and
\ref{fig:prob1}.  In all cases $\chi^2$ decreased at least eight orders of
magnitude from its initial value during the simulated annealing calculation.

\begin{figure}[ht]
 \centering
 \includegraphics[width=\linewidth]{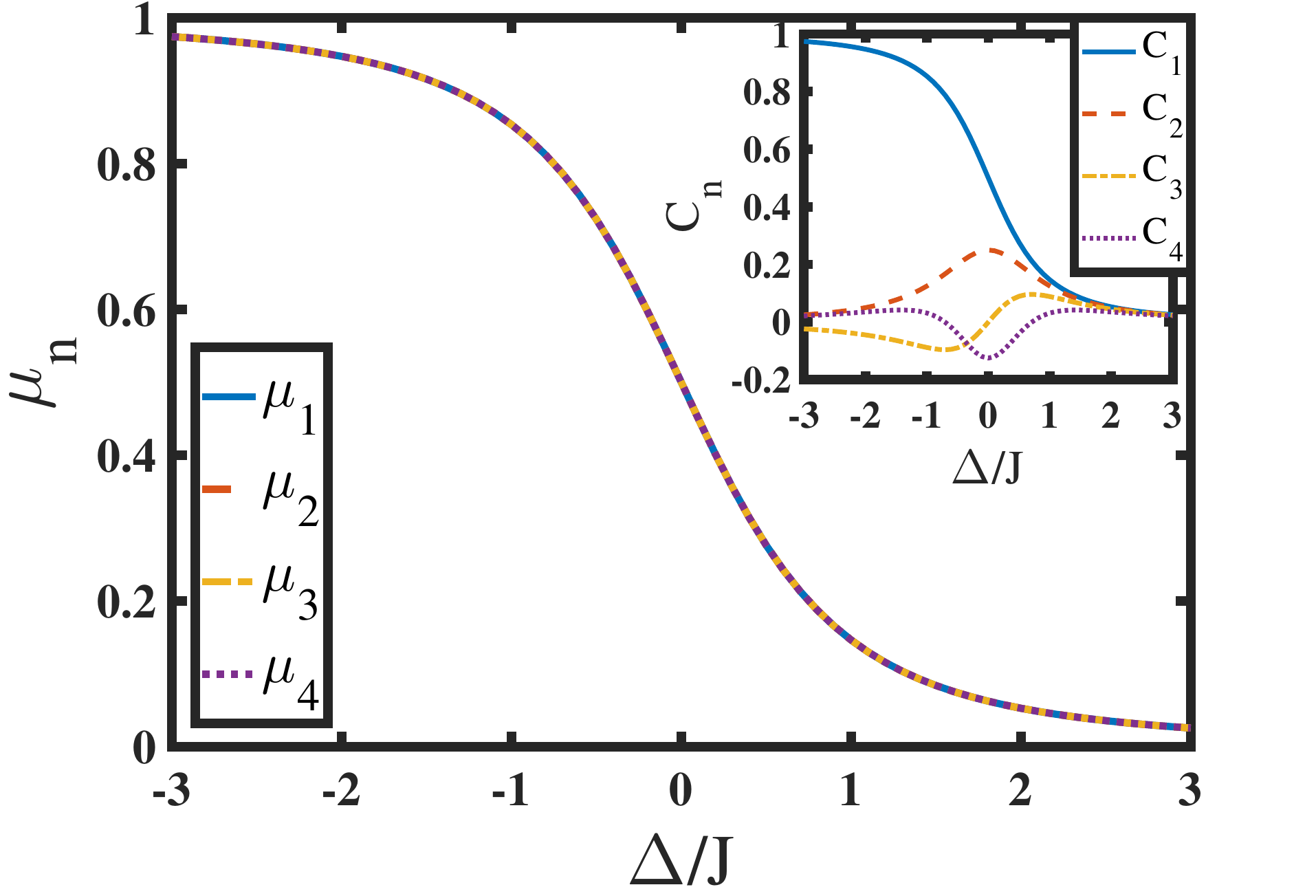}
 \includegraphics[width=\linewidth]{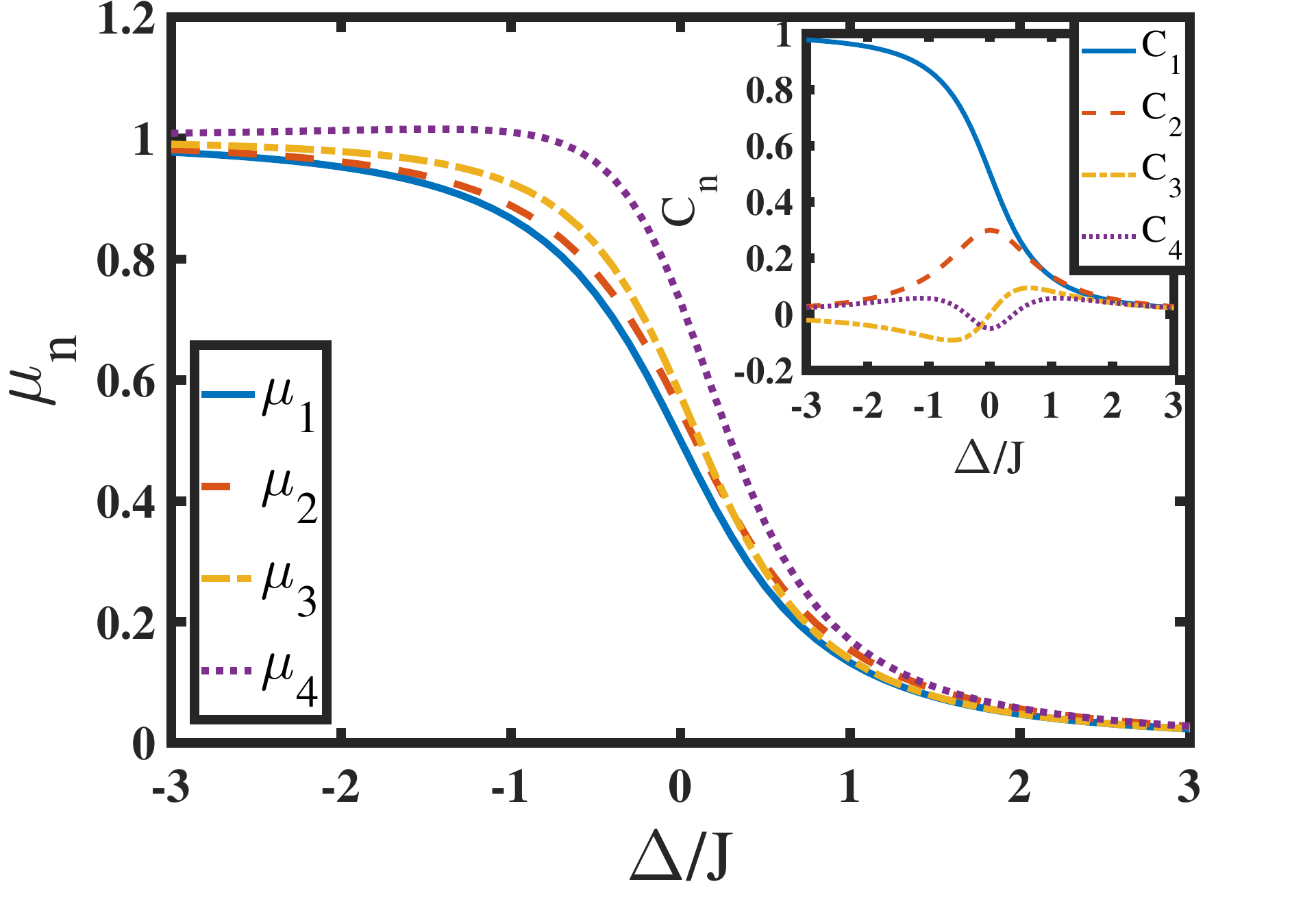}
 \includegraphics[width=\linewidth]{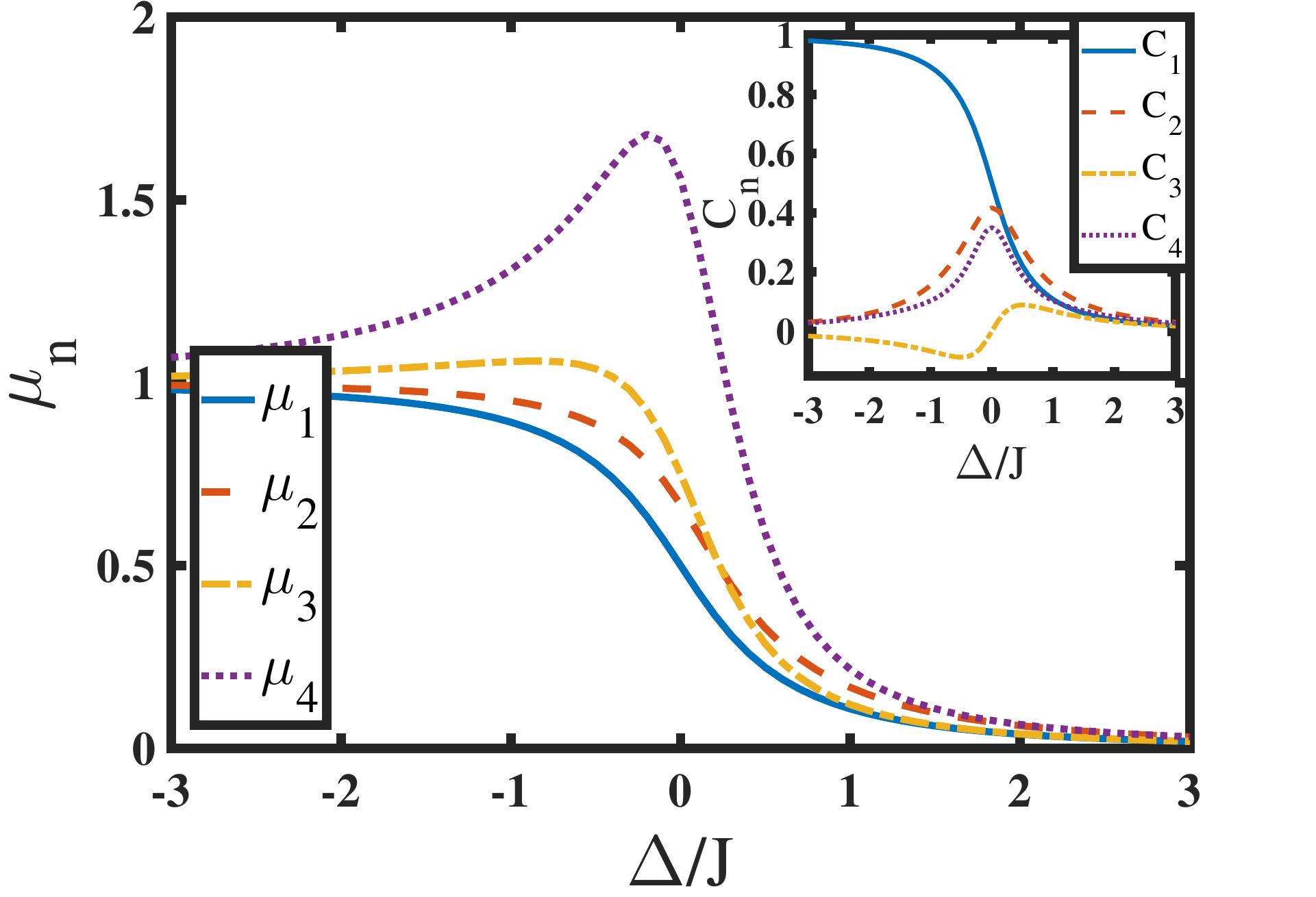}
 \includegraphics[width=\linewidth]{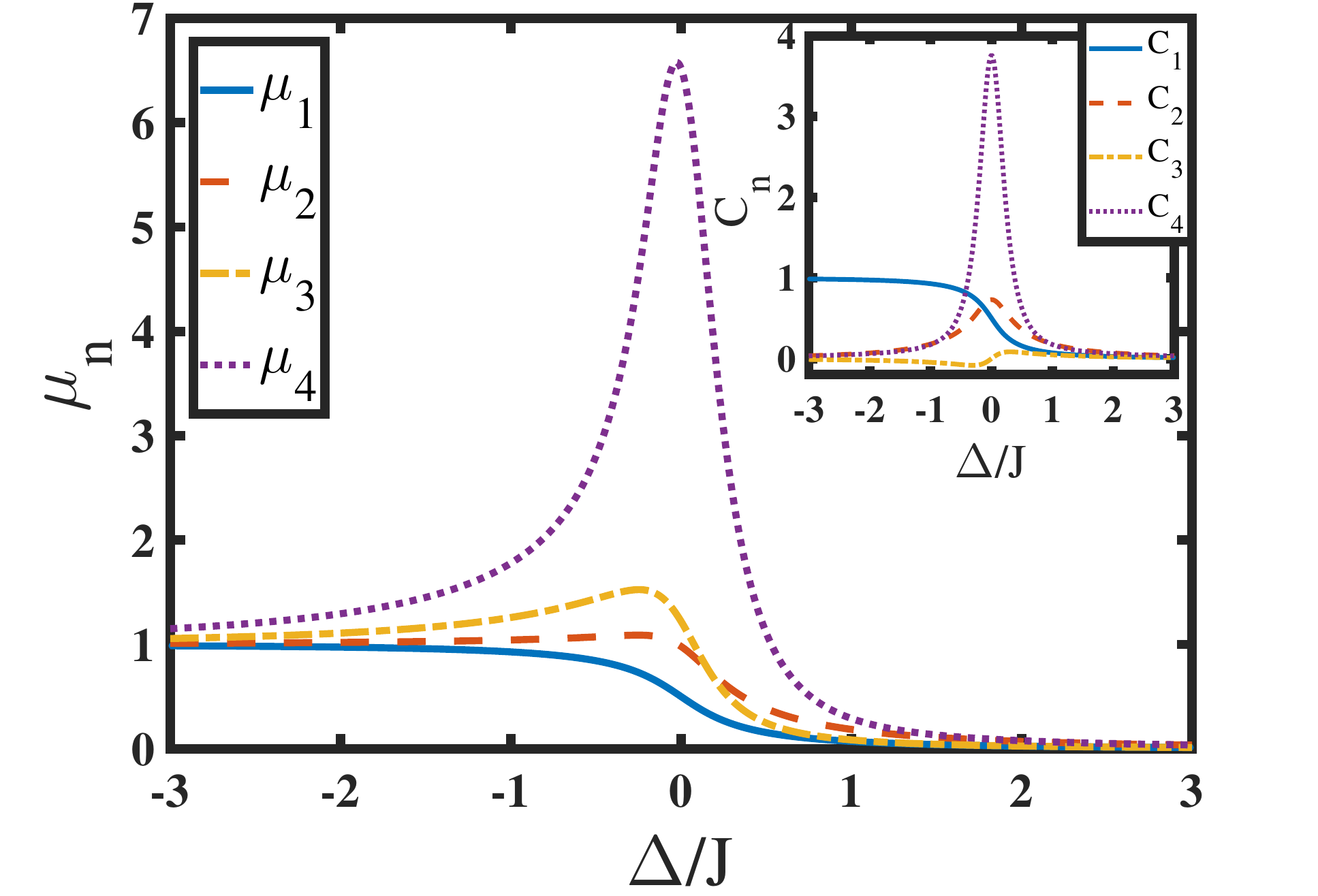}
 \caption{Moments for $J'/J=0, 0.3, 0.5, 0.7$ as a function of $\Delta/J$.
   In these calculations $J=1$. For $J'/J=0$ the curves are
   identical.  The insets show the corresponding cumulants.}
 \label{fig:mom01}
\end{figure}

\begin{figure}[ht]
 \centering
 \includegraphics[width=\linewidth]{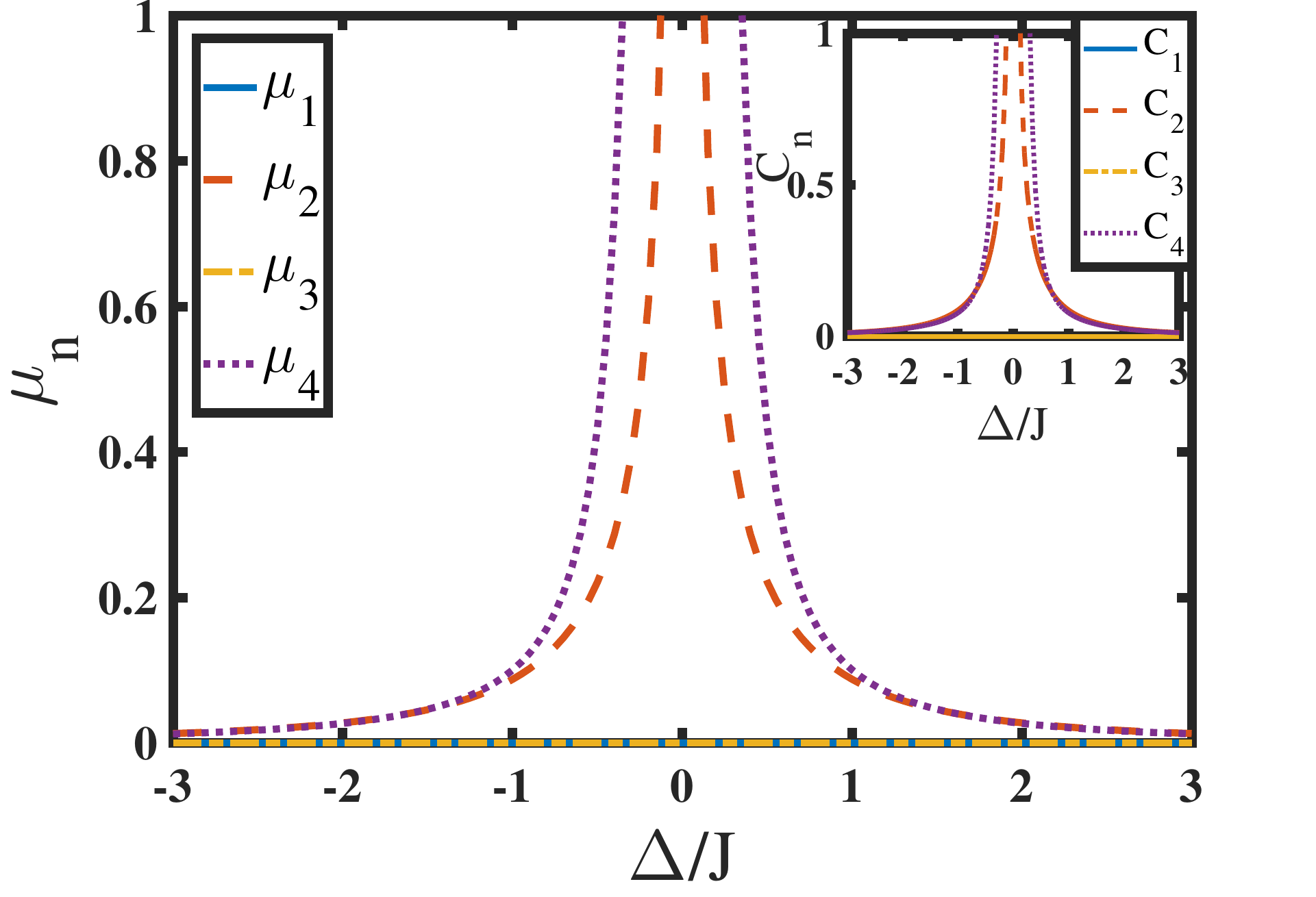}
 \includegraphics[width=\linewidth]{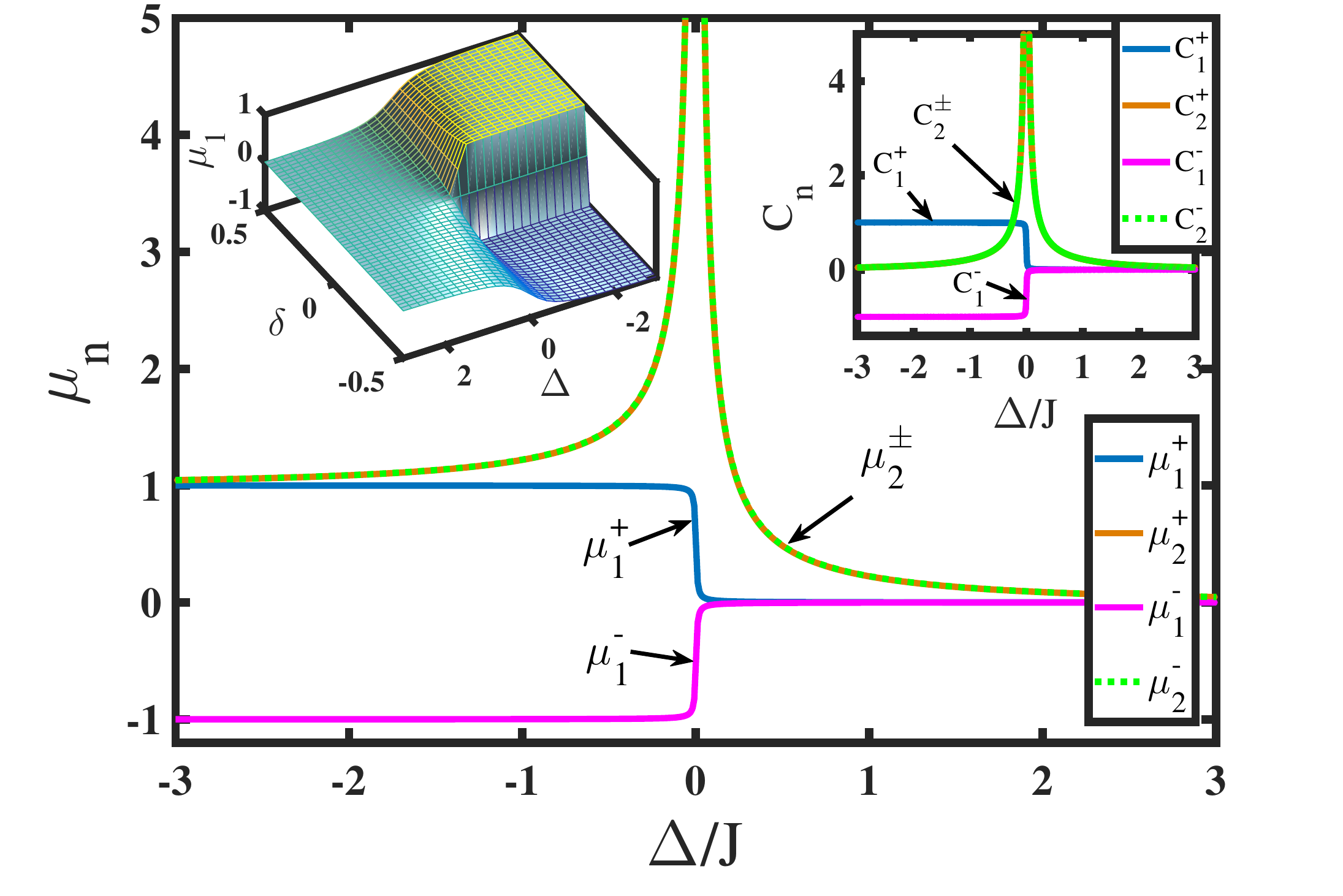}
 \caption{(Color online.)  Upper panel: Moments for $J'/J=1$ as a function of
   $\Delta/J$.  In these calculations $J=1$. The inset shows the corresponding
   cumulants.  In the limit $\Delta/J \rightarrow 0$ (the topological point of
   the model) the even cumulants diverge, while the odd cumulants are always
   zero for this case.  Lower panel: first two moments and cumulants (right
   inset) for $J=1,J'=J \pm 0.006$.  Left inset shows the first moment on the
   $\Delta-\delta$ plane, indicating the singular behavior along the line
   $\Delta < 0, \delta = 0$.}
 \label{fig:mom1}
\end{figure}

\begin{figure*}[b]
\includegraphics[width=\linewidth]{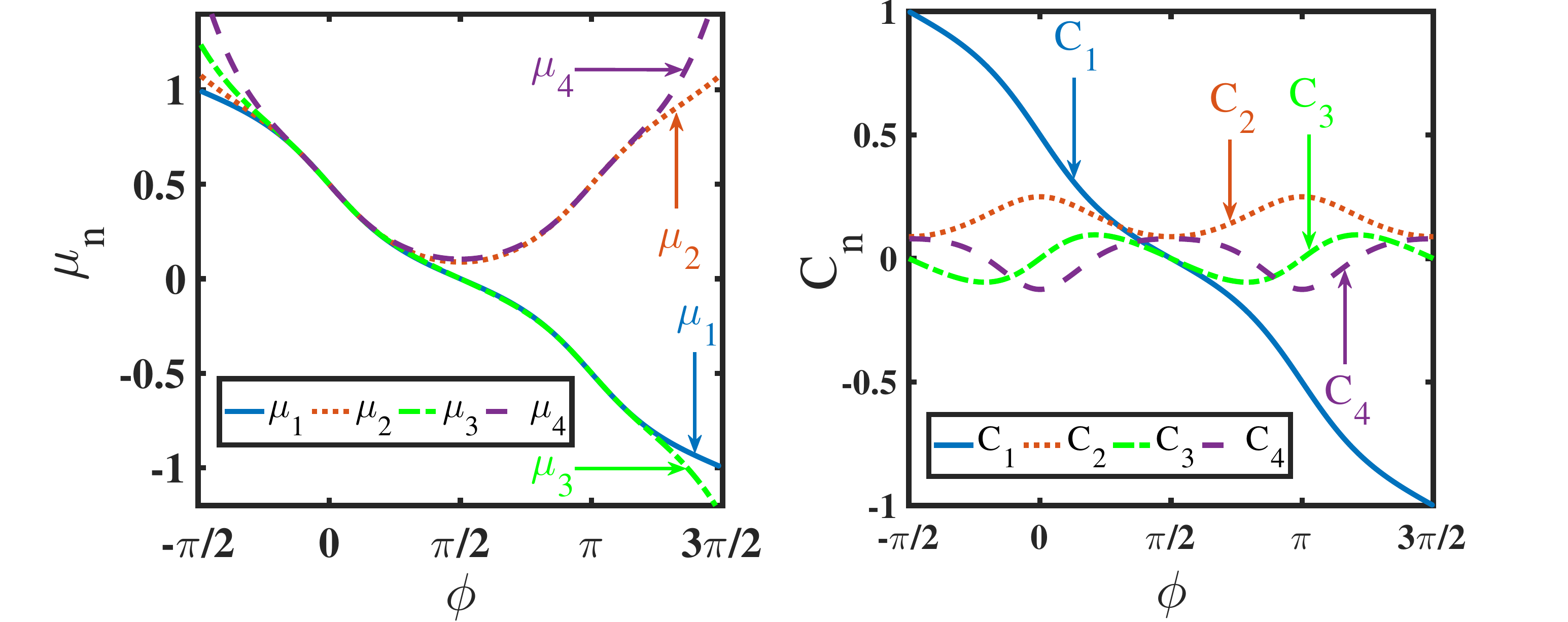}
\includegraphics[width=\linewidth]{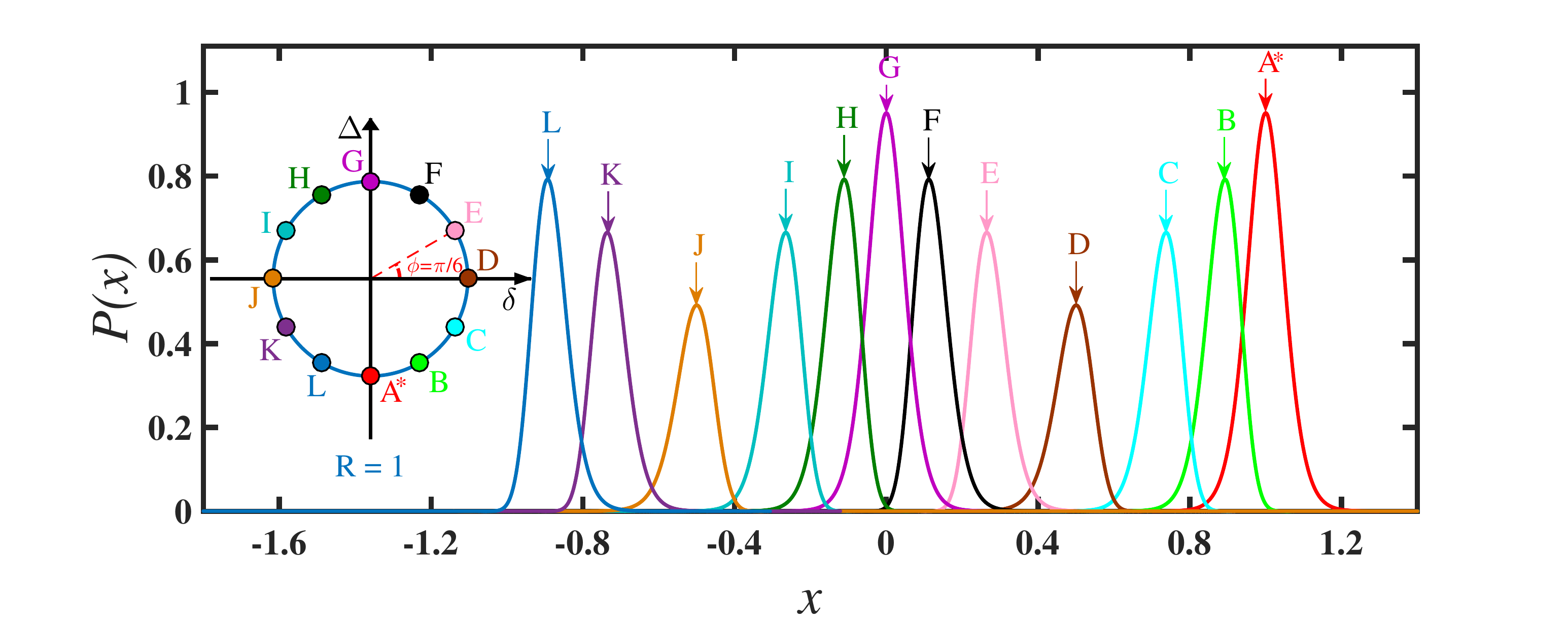}
\caption{(Color online.) Moments, cumulants, and probability distribution
  along a circle of radius $1$ in the $\Delta/t$ vs. $\delta/t$ plane.  In
  these calculations $t=1$. The path encircles the topological point
  $\Delta/t=0,\delta/t=0$.  The upper panel shows the gauge invariant moments
  and cumulants along the circle as a function of angle.  The lower panel
  follows the evolution of the probability distribution.  The point $A^*$ is
  at an angle $\phi = -\pi/2 + 2 \pi/1000$, not $\phi = -\pi/2$.  In the lower
  panel the unit of $x$ is the lattice constant.  The points $\phi =
  -\pi/2, 3\pi/2$ are excluded from the curves shown in the upper panels. }
\label{fig:CircleR1}
\end{figure*}

\begin{figure*}[b]
\includegraphics[width=\linewidth]{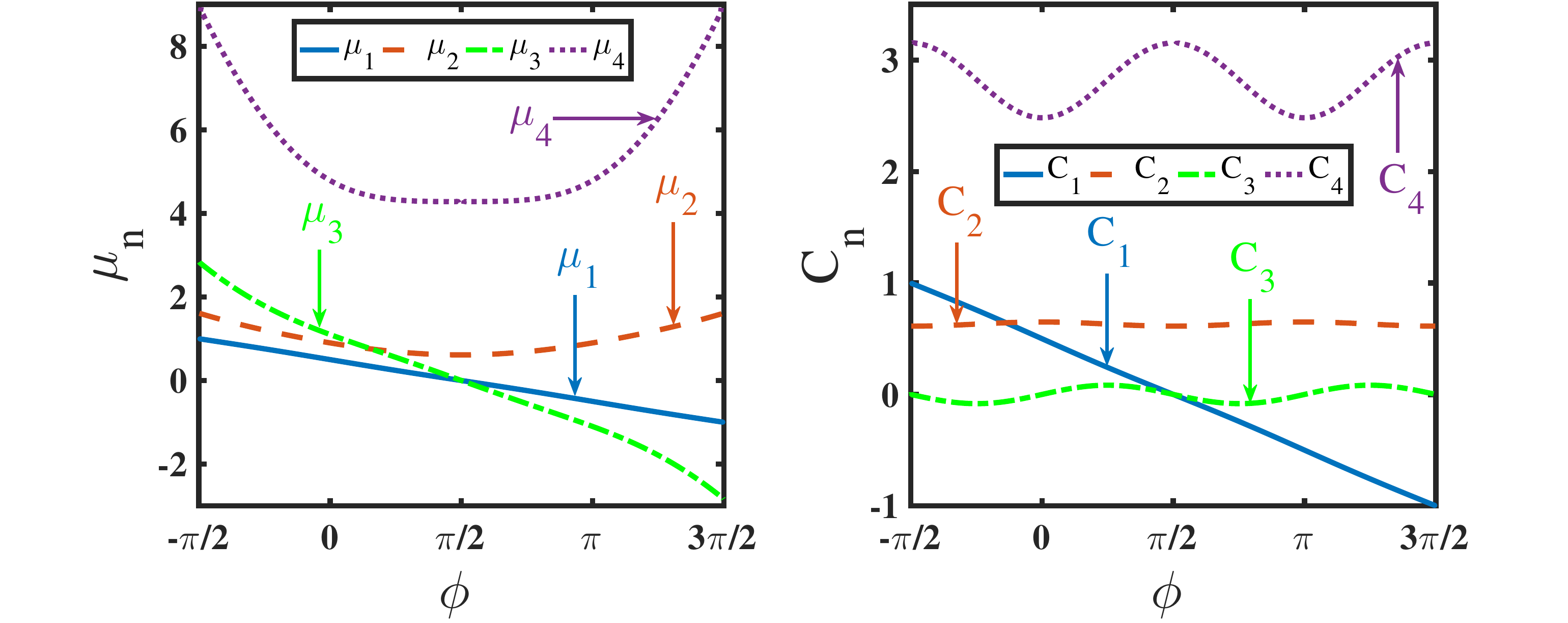}
\includegraphics[width=\linewidth]{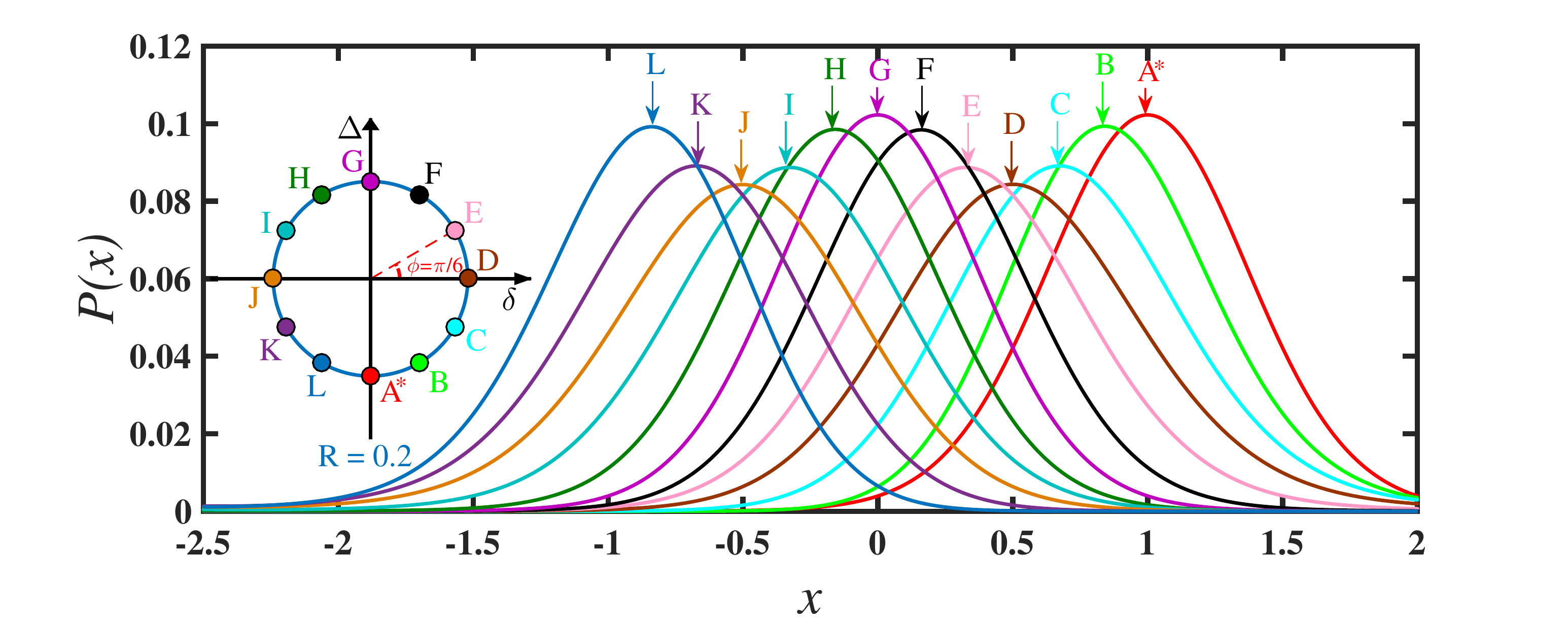}
\caption{(Color online.) Moments, cumulants, and probability distribution
  along a circle of radius $0.2$ in the $\Delta/t$ vs. $\delta/t$ plane.  In
  these calculations $t=1$.  The path encircles the topological point
  $\Delta/t=0,\delta/t=0$.  The upper panel shows the gauge invariant moments
  and cumulants along the circle as a function of angle.  The lower panel
  follows the evolution of the probability distribution. The point $A^*$ is at
  an angle $\phi = -\pi/2 + 2 \pi/1000$, not $\phi = -\pi/2$.  In the lower
  panel the unit of $x$ is the lattice constant.  The points $\phi =
  -\pi/2, 3\pi/2$ are excluded from the curves shown in the upper panels. }
\label{fig:CircleR02}
\end{figure*}

In Table \ref{tab:chi2_circles} the values of the parameters according to the
parametrization used in Figs. \ref{fig:mom01}-\ref{fig:prob1} are shown.  Also
shown are values of the negative base ten logarithm of $\chi^2$ rounded down
to the first digit for probability distributions corresponding to the points
in Figs. \ref{fig:CircleR1} and \ref{fig:CircleR02}.

\clearpage
\newpage

\begin{table}
\begin{tabular}{|c||c|c||c|c|}
\hline
Fig. \ref{fig:prob} panel 1 & $\Delta/J$ & $-\log_{10}(\chi^2)$ & $\Delta/J$ & $-\log_{10}(\chi^2)$  \\ \hline \hline
$J'/J = 0$ & $-2$ &  $8$ & $-1$ &  $7$ \\ \hline 
$J'/J = 0$ & $-0.5$ & $8$  & $-0.4$ & $6$  \\ \hline 
$J'/J = 0$ &$-0.3$ & $7$  & $-0.2$ &  $6$ \\ \hline 
$J'/J = 0$ &$-0.1$ & $7$ & $0$ &  $6$ \\ \hline \hline
Fig. \ref{fig:prob} panel 2 & $\Delta/J$ & $-\log_{10}(\chi^2)$ & $\Delta/J$ & $-\log_{10}(\chi^2)$  \\ \hline \hline
$J'/J = 0.3$ & $-2$ &  $6$ & $-1$ &  $5$ \\ \hline 
$J'/J = 0.3$ & $-0.5$ & $6$  & $-0.4$ & $4$  \\ \hline 
$J'/J = 0.3$ &$-0.3$ & $5$  & $-0.2$ &  $4$ \\ \hline 
$J'/J = 0.3$ &$-0.1$ & $5$ & $0$ &  $4$ \\ \hline \hline
Fig. \ref{fig:prob} panel 3 & $\Delta/J$ & $-\log_{10}(\chi^2)$ & $\Delta/J$ & $-\log_{10}(\chi^2)$  \\ \hline \hline
$J'/J = 0.5$ & $-2$ &  $5$ & $-1$ &  $4$ \\ \hline 
$J'/J = 0.5$ & $-0.5$ & $5$  & $-0.4$ & $4$  \\ \hline 
$J'/J = 0.5$ &$-0.3$ & $5$  & $-0.2$ &  $4$ \\ \hline 
$J'/J = 0.5$ &$-0.1$ & $5$ & $0$ &  $4$ \\ \hline \hline
Fig. \ref{fig:prob} panel 4 & $\Delta/J$ & $-\log_{10}(\chi^2)$ & $\Delta/J$ & $-\log_{10}(\chi^2)$  \\ \hline \hline
$J'/J = 0.7$ & $-2$ &  $5$ & $-1$ &  $4$ \\ \hline 
$J'/J = 0.7$ & $-0.5$ & $5$  & $-0.4$ & $4$  \\ \hline 
$J'/J = 0.7$ &$-0.3$ & $4$  & $-0.2$ &  $4$ \\ \hline 
$J'/J = 0.7$ &$-0.1$ & $4$ & $0$ &  $4$ \\ \hline \hline
Fig. \ref{fig:prob1} & $\Delta/J$ & $-\log_{10}(\chi^2)$ & $\Delta/J$ & $-\log_{10}(\chi^2)$  \\ \hline \hline
$J'/J = 1.0$ & $-1$ &  $8$ & $-0.6$ &  $7$ \\ \hline 
$J'/J = 1.0$ & $-0.9$ & $8$  & $-0.5$ & $7$  \\ \hline 
$J'/J = 1.0$ &$-0.8$ & $7$  & $-0.4$ &  $7$ \\ \hline 
$J'/J = 1.0$ &$-0.7$ & $7$ & &   \\ \hline \hline
\end{tabular}
\caption{$-\log_{10}\chi^2$ rounded to the first digit shown for the
  reconstructed probabilities in Figs. \ref{fig:prob} and \ref{fig:prob1}.  }
\label{tab:chi2}
\end{table}           

\begin{table}
\begin{tabular}{|c||c|c|c||c|c|c|}
\hline 
 & \multicolumn{3}{c||}{  Fig. \ref{fig:CircleR1} } & \multicolumn{3}{|c|}{  Fig. \ref{fig:CircleR02} } \\
\hline \hline 
& $\Delta/J$ & $J'/J$ & $-\log_{10}(\chi^2)$ &
 $\Delta/J$ & $J'/J$ & $-\log_{10}(\chi^2)$  \\ \hline \hline
$A^*$ & $-1.9875$ & $0.9875$ & $8$ &
 $-.3995$ &  $.9975$ & $6$  \\ \hline
$B$ &  $-1.154$ & $.333$ &  $7$ & 
  $-.3149$& $.8181$ &  $6$ \\ \hline
$C$ &  $-.5359$ & $.072$ &  $7$ & 
  $-.1705$ & $.7047$ &  $6$ \\ \hline
$D$ &  $0$ & $0$ &  $6$ & 
 $0$ & $.6666$ &  $6$ \\ \hline
$E$ &  $.5359$ &  $.072$ &  $8$ & 
 $.1705$ & $.7047$ &  $6$ \\ \hline
$F$ &  $1.154$ &  $.333$ &  $7$ & 
  $.3149$& $.8181$ &  $6$ \\ \hline
$G$ &  $2$ &  $1$ &  $7$ & 
  $.4$& $1$ &  $6$ \\ \hline
$H$ &  $3.4641$ & $3$ &  $6$ & 
  $.3849$ & $1.2222$ &  $6$ \\ \hline
$I$ &  $7.4641$ & $13.928$ &  $8$ & 
  $.2419$ & $1.4189$ &  $6$ \\ \hline
$J$ &  $0$ &  $\infty$ $(1/0)$ &  $7$ & 
  $0$  & $1.5$  &  $6$ \\ \hline
$K$ &  $-7.4641$ & $13.928$ &  $7$ & 
  $-.2419$ & $1.4189$ &  $6$ \\ \hline
$L$ &  $-3.4641$ & $3$ &  $6$ & 
  $-.3849$& $1.2222$ &  $6$ \\ \hline
\end{tabular}
\caption{Values of the parameters according to the parametrization used in
  Figs. \ref{fig:mom01}-\ref{fig:prob1} are shown. Also shown are values of
  $-\log_{10}\chi^2$ rounded to the first digit for probability distributions
  corresponding to the points in Figs. \ref{fig:CircleR1} and
  \ref{fig:CircleR02}. }
\label{tab:chi2_circles}
\end{table}

\end{document}